\documentclass[twocolumn]{jpsj3} 
\usepackage{color}

\title{Diamagnetism above $T_{\rm c}$ in underdoped ${\rm Bi}_{2.2}{\rm Sr}_{1.8}{\rm Ca}_{2}{\rm Cu}_{3}{\rm O}_{10+\delta}$}

\author{Tetsuya \textsc{Iye}$^{1}$\thanks{E-mail: tiye@scphys.kyoto-u.ac.jp}\thanks{Present address:  Department of Physics, Graduate School of Science, Kyoto University, Kyoto 606-8502.}, Takenori \textsc{Nagatochi}$^{1}$\thanks{Present address:  Department of Advanced Materials Science, Graduate School of Frontier Science, the University of Tokyo, Chiba 277-8561.}, Ryusuke \textsc{Ikeda}$^{2}$, and Azusa \textsc{Matsuda}$^{1}$}

\inst{$^{1}$Department of Physics, School of Science and Engineering, Waseda University, Shinjuku, Tokyo 169-8555 \\
	$^{2}$Department of Physics, Graduate School of Science, Kyoto University, Kyoto 606-8502}

\abst{Single crystals of ${\rm Bi}_{2+x}{\rm Sr}_{2-x}{\rm Ca}_{2}{\rm Cu}_{3}{\rm O}_{10+\delta}$(Bi2223) with $x=0.2$ were grown by a traveling solvent floating zone method in order to investigate the superconducting properties of highly underdoped Bi2223.
Grown crystals were characterized by X-ray diffraction, DC susceptibility and resistivity measurements, confirming Bi2223 to be the main phase.
The crystals were annealed under various oxygen partial pressures to adjust their carrier densities from optimally doped to highly underdoped.
The fluctuation diamagnetic component above the superconducting transition temperature $T_{\rm c}$ extracted from the anisotropic normal state susceptibilities $\chi_{ab}(T)$ ($H\perp c$) and $\chi_{c}(T)$ ($H\parallel c$) was found to increase with underdoping, suggesting a decrease in the superconducting dimensionality and/or increase in the fluctuating vortex liquid region.}

\kword{high-$T_{\rm c}$ cuprate, Bi2223, single crystal, normal state susceptibility, dimensionality}

\begin{document}
\maketitle

\section{Introduction}
In multilayered cuprate superconductors, which have more than three $\rm CuO_{2}$ layers in their unit cells, the superconducting transition temperature $T_{\rm c}$ takes its highest value when the number of $\rm CuO_{2}$ layers $n$ is 3 \cite{Iyo2007, Chakravarty2004}.
The $T_{\rm c}$ reduction for increasing $n$ can be understood in terms of inequivalent doping among the layers.
However, the mechanism which enhances the $T_{\rm c}$ for the first few layers is still unknown and may be related to the mechanism of high-$T_{\rm c}$ superconductivity.

Nuclear magnetic resonance (NMR), and angle resolved photoemission spectroscopy (ARPES) have shown evidence of the inequivalent doping among layers.
The inner $\rm CuO_{2}$ planes (IPs) with square oxygen coordination have lower hole doping levels than the outer $\rm CuO_{2}$ planes (OPs) with pyramidal oxygen coordination \cite{Trokiner1991, Statt1993, Kontos1998, Piskunov1998, Tokunaga1999, Kotegawa2001, Ideta2009, Chen2009}.
In the NMR experiments, the remarkable coexistence of antiferromagnetic (AFM) IPs with superconducting (SC) OPs has been reported in the underdoped regime \cite{Kotegawa2004, Mukuda2006, Kitaoka2007, Shimizu2009}.
It has been shown theoretically that high $T_{\rm c}$ superconductivity is possible in this AFM and SC coexisting state \cite{Mori2005}.
In that case, a finite Josephson coupling between SC OPs through the AFM IP provides bulk superconductivity. 
Thus, physical properties of underdoped region of multilayered superconductor is intriguing.

In the overdoped to slightly underdoped region, several anisotropic properties of a trilayered superconductor ${\rm Bi}_{2}{\rm Sr}_{2}{\rm Ca}_{2}{\rm Cu}_{3}{\rm O}_{10+\delta}$ (Bi2223) have been measured and a sizable change in anisotropy was reported \cite{Piriou2008}.
The anisotropy was lower than that of double layered superconductor ${\rm Bi}_{2}{\rm Sr}_{2}{\rm Ca}{\rm Cu}_{2}{\rm O}_{10+\delta}$ (Bi2212), indicating that the interlayer coupling of the three $\rm CuO_{2}$ layers exists.
Bi2223 ($T_{\rm c,max}$ = 110 K) is one of the few multilayered cuprates of which we can obtain large size single crystals in order of millimeters \cite{Fujii2001, Liang2002, Giannini2004, Tokiwa1998, Iyo2004} and is the simplest system among multilayered cuprates.
In this paper, we first report on a growth of Bi2223 crystals suitable for underdoping.
Then, the doping dependence of fluctuation diamagnetism in Bi2223 is extracted from magnetization measurements.
The extracted diamagnetism above $T_{\rm c}$ is expected to be qualitatively the same as the diamagnetism which Li \textit{et al} observed through Nernst and torque magnetization experiment.\cite{Li2010}

\section{Experimental}
Two techniques were used to reduce the doping level and drive ${\rm Bi}_{2+x}{\rm Sr}_{2-x}{\rm Ca}_{2}{\rm Cu}_{3}{\rm O}_{10+\delta}$ into the highly underdoped regime.
The first is to anneal samples under various oxygen partial pressures.
Oxygen annealing controls the excess oxygen content $\delta$, which supplies hole carriers to the $\rm CuO_{2}$ plane.
The hole doping level can be decreased by lowering the oxygen partial pressure \cite{Watanabe1997}.
The second method is a substitution of the trivalent cation $\rm Bi^{3+}$ for the divalent cation $\rm Sr^{2+}$.
The hole density is expected to decrease with Bi substitution if the oxygen content is unchanged.
In Bi2212, a hole density reduction was confirmed through $T_{\rm c}$ reduction due to this substitution \cite{Yamashita2009}.
${\rm Bi}_{2+x}{\rm Sr}_{2-x}{\rm Ca}_{2}{\rm Cu}_{3}{\rm O}_{10+\delta}$ single crystals with $x=0.1$ can be grown by a traveling solvent floating zone (TSFZ) method \cite{Kulakov2006, Fujii2002, Eisaki2004}.
However, no successful single crystal growth of Bi2223 with $x>0.1$ has been reported, because the growth is difficult even for the intrinsically substituted $x=0.1$ sample.
Here, using almost the same technique as for $x=0.1$, we succeeded in growing crystals of Bi2223 with $x=0.2$.

\begin{figure}[t]
	\begin{center}
	\begin{tabular}{cc}
		\begin{minipage}{0.682\hsize}
			\begin{center}
			\includegraphics[width=5.797cm,clip]{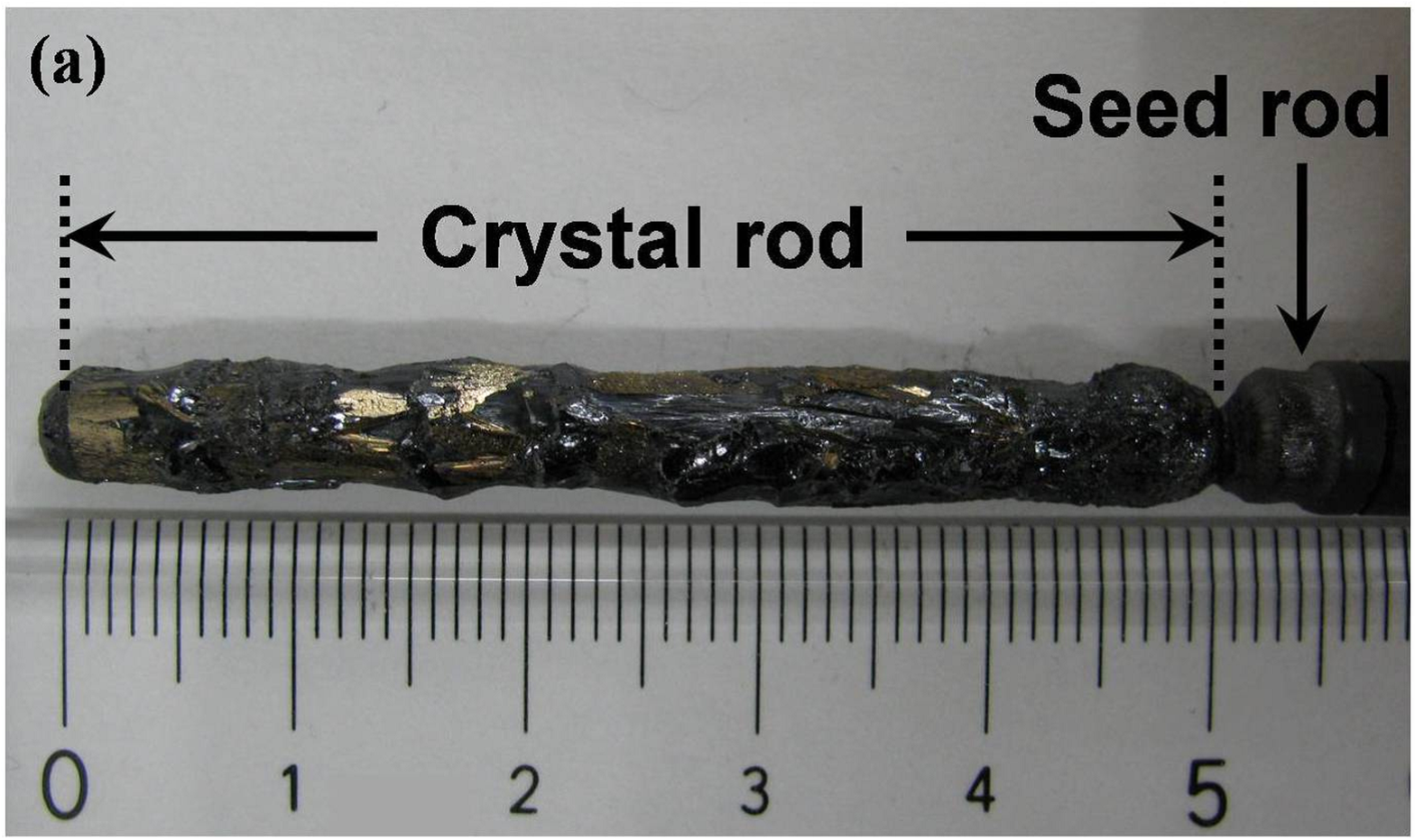}
			\end{center}
		\end{minipage}
		\begin{minipage}{0.318\hsize}
			\begin{center}
			\includegraphics[width=2.703cm,clip]{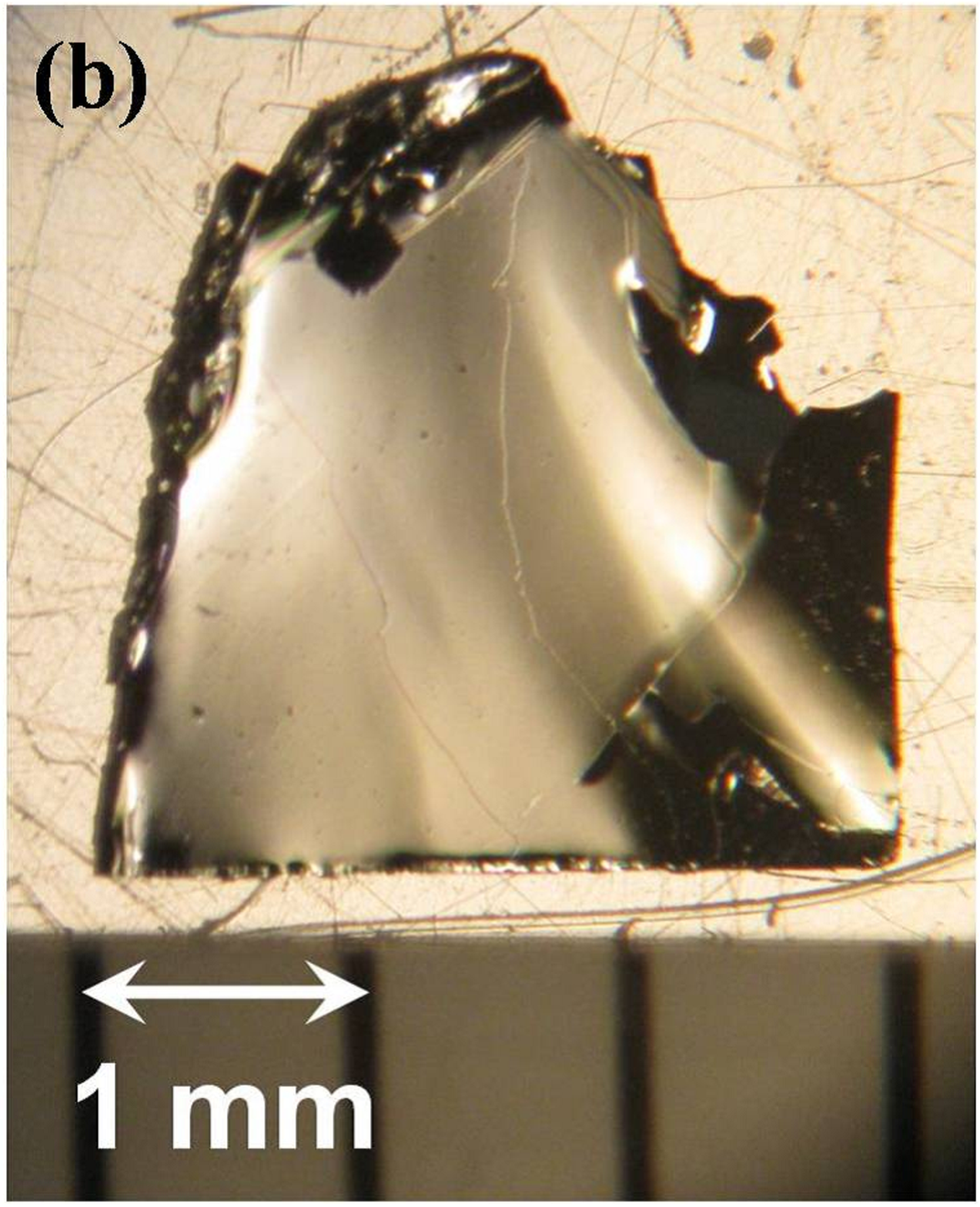}
			\end{center}
		\end{minipage}
		\end{tabular}
		\caption{(Color online) (a)The grown crystal rod of 47 mm in length and typical diameter 5.5 mm. The linear scale is in centimeters. (b)One of the Bi2223 single crystals cleaved from the rod shown in (a).}{}
		\label{fig:crystal}
	\end{center}
\end{figure}

Powders of $\rm Bi_{2}O_{3}$, $\rm SrCO_{3}$, $\rm CaCO_{3}$, and CuO (all of 99.9 \% or higher purity) were mixed in the desired cation ratio Bi : Sr : Ca : Cu = 2.2 : 1.8 : 2 : 3, ground in an agate mortar, and then calcined at 760 $^\circ\mathrm{C}$ for 12 hours.
The calcined powders were well reground and again calcined at 780 $^\circ\mathrm{C}$ for 12 hours.
After that, the powders were hydrostatically pressed into a cylindrical rod under 40 MPa and then sintered at 860 $^\circ\mathrm{C}$ for 50 hours.
The sintered rod was cut into the long ($\sim7$ cm) and the short ($\sim2$ cm) part.
The long one was hung on the upper shaft of an infrared radiation furnace equipped with two ellipsoidal mirrors (NEC-SCI\hspace{-.1em}I-EDH).
The short one was held on the lower shaft.
The long one was pre-melted using the short one as a base for the crystal growth at a rate of 25 mm/h in air.
The pre-melted rod was cut into the pre-melted part ($\sim 6$ cm) and the base (seed) part ($\sim 2$ cm).
The pre-melted and the seed rod were held on the upper and the lower shafts respectively, and both shafts were counter-rotated at a rate of 11 rpm.
A crystal rod was grown using a very slow rate of 0.04 mm/h for the pre-melted rod.
In order to realize as fast a crystal growth rate as possible, 300 W halogen lamps were used as light sources, which gave a steep temperature gradient near the solid-liquid interface.
It also prevents the swells of solid-liquid interfaces on both the pre-melted and grown crystal sides which disturb static growth.
Under these conditions, a crystal rod of 47 mm length and 5.5 mm in diameter as shown in Fig. \ref{fig:crystal}(a) was grown.

From the grown crystal rod, one crystal with maximum dimensions up to $\rm 2 mm \times 3 mm \times 0.1 mm$ was cleaved and is shown in Fig. \ref{fig:crystal}(b).
Each part of the as-grown rod was evaluated by X-ray diffraction (XRD), magnetizaton and resistivity measurements.
XRD measurements were done on a diffractometer (Rigaku  RINT-TTRI\hspace{-.1em}I\hspace{-.1em}I) using a conventional $\theta-2\theta$ method to check phase purity and determine $c$-axis length.
$T_{\rm c}$ was determined from magnetization measurements under a magnetic field of 4 Oe using a SQUID magnetometer (Quantum Design MPMS-7).
Resistivity was measured by a standard four-probe method.
After finishing the above sample characterizations, the highest quality crystal among them, whose weight was 8.6 mg, was selected.
After each annealing condition; (1) $\rm O_{2}$ : 760 torr, 400 $^\circ\mathrm{C}$, 72 hours, (2) $\rm O_{2}$ : 760 torr, 600 $^\circ\mathrm{C}$, 24 hours, (3) $\rm O_{2}$ : 7.6 torr, 600 $^\circ\mathrm{C}$, 24 hours, (4) $\rm O_{2}$ : $7.6\times10^{-2}$ torr, 600 $^\circ\mathrm{C}$, 24 hours, (5) $\rm O_{2}$ : $7.6\times10^{-3}$ torr, 600 $^\circ\mathrm{C}$, 12 hours, $T_{\rm c}$ and $c$-axis length were determined, then anisotropic normal state susceptibilities $\chi_{ab}$ with $H\perp c$ and $\chi_{c}$ with $H\parallel c$ were measured under a magnetic field of 50 kOe using the MPMS.
For each annealing process, 1 hour was taken to raise temperature to the target value, while oxygen pressure was kept at the target value.
At the end of each annealing step, the sample was quenched to room temperature in 20 minutes in the same atmosphere in order to prevent a further oxygen uptake or loss.
SCF diamagnetic components were extracted from the normal state susceptibility data for each annealing condition.

\section{Results and Discussion}

\begin{figure}[t]
	\begin{center}
	\includegraphics[width=8.5cm,clip]{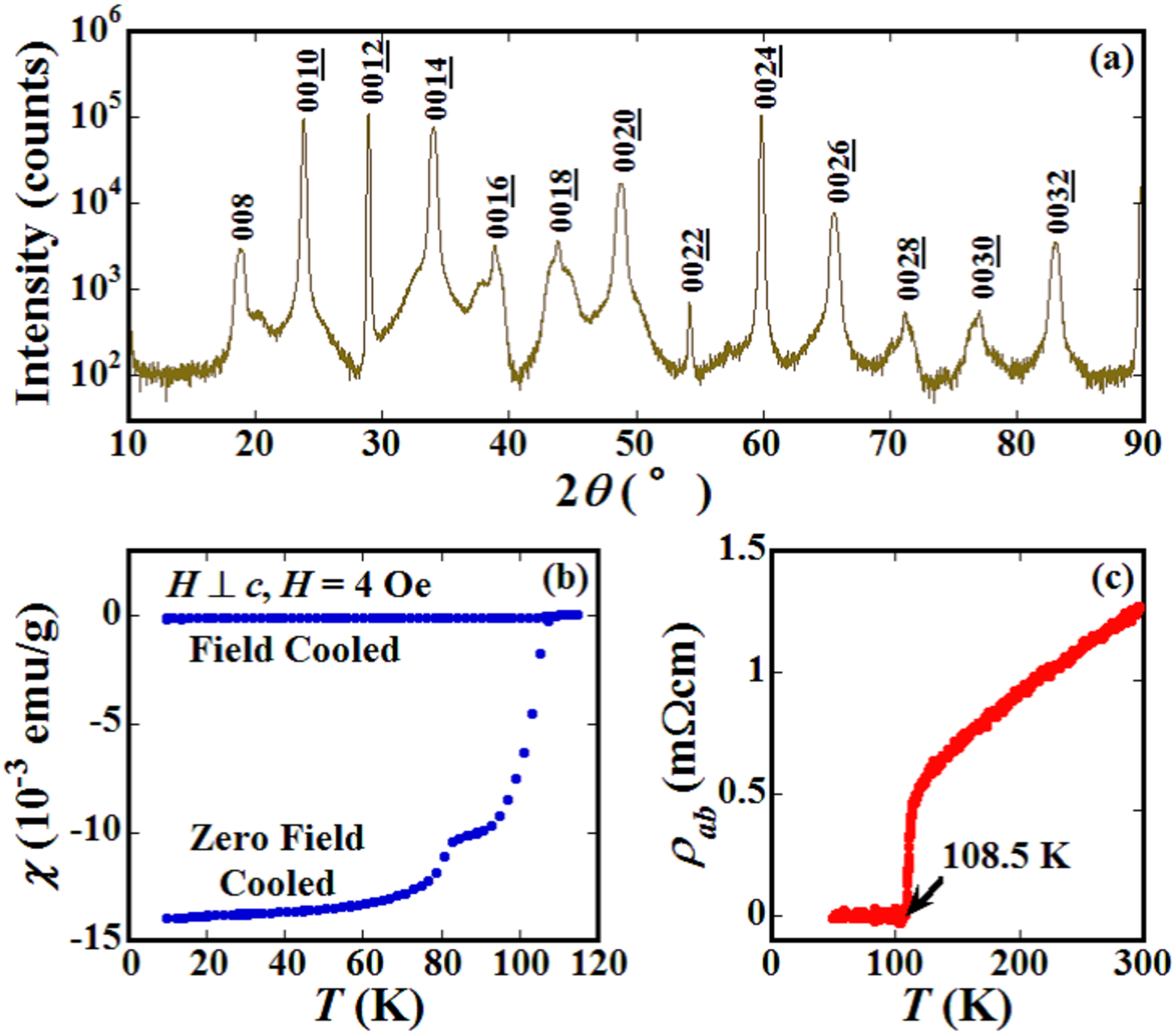}
 	\caption{(Color online) (a)X-ray diffraction peaks of the as-grown sample with $x=0.2$. (b)Temperature dependence of DC susceptibility for the same sample with magnetic field 4 Oe parallel to $ab$-axis. (c)Temperature dependence of in-plane resistivity for another as-grown sample.}{}
 	\label{fig:XRD-sus-res}
	\end{center}
\end{figure}

Figure \ref{fig:XRD-sus-res}(a) shows XRD peaks in the highest quality $c$-axis oriented as-grown crystal.
Only reflections from Bi2223 can be seen.
Figure \ref{fig:XRD-sus-res}(b) shows the temperature dependence of the magnetic susceptibility in a magnetic field of 4 Oe parallel to the $ab$-plane.
Clear Meissner diamagnetism was observed below about 110 K.
Throughout this report, $T_{\rm c}$s are defined as the temperatures at which linear extrapolations of the steepest part of shielding signals cross zero susceptibility.
Using this definition, $T_{\rm c}$ of the main phase was estimated to be 110 K.
A step-like feature at around 80 K which may attributed to Bi2212 impurity phase was also observed in the shielding signal.
Since XRD did not show any traces of bulk Bi2212 phase, it might be due to Bi2212 intergrowth, existing even in the highest quality crystal.
By comparing the magnitude of the susceptibility at 80 K with the magnitude at 10 K, the relative volume fraction of Bi2223 phase to that of overall superconducting phase was estimated to be more than 60 \%.
The temperature dependence of the in-plane resistivity of a crystal cleaved from the same part of the rod was shown in Fig. \ref{fig:XRD-sus-res}.
It showed typical $T$-linear behavior above $T_{\rm c}$ and zero resistivity below 108.5 K.

\begin{figure}[t]
	\begin{center}
	\includegraphics[width=8.5cm,clip]{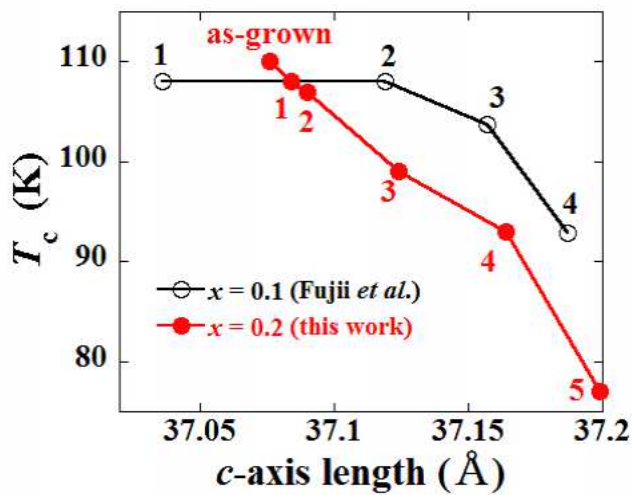}
 	\caption{(Color online) Comparison of the relationships between $T_{c}$ and $c$-axis length for $x=0.2$ (filled circle) and $x=0.1$ (open circle). The attached numbers correspond to the annealing conditions (1) to (5) for both $x=0.1$ and 0.2 data. Data for $x=0.1$ were taken from ref.\cite{Fujii2002}.}{}
 	\label{fig:c-Tc}
	\end{center}
\end{figure}

Next, let us discuss the precise determination of the $c$-axis lengths in the Bi2223 crystal for $x=0.2$.
The $c$-axis length free from systematic measurement errors is usually estimated by an extraporation method using a Nelson-Riley (NR) function $(\cos^{2}\theta/\sin\theta+\cos^{2}\theta/\theta)/2$.
If $c$-axis lengths obtained from (00$L$) peaks \textit{i.e.} $dL$, where $d = \lambda/2\sin\theta_{L}$, $L$ is the Miller index of the (001) direction are plotted against the NR function and the plots are fit with a straight line, an accurate $c$-axis length can be obtaind by extrapolating the fitting line to $\theta = \pi/2$.
However, the plots for our crystal with $x=0.2$ did not obey this function.
It is known that if there is intergrowth of a second phase in a layered structure, $dL$ can oscillate as a function of $L$ with a period $c/\Delta c$, where $\Delta c$ is the difference between the $c$-axis lattice length of the main phase and that of the intergrown phase \cite{Kulakov2006}.
However, due to the strong and somewhat diverging oscillation in the plots of our sample, it was difficult to execute a reliable extrapolation.
Here, we computed $c$-axis lengths for $x=0.2$ by averaging $dL$ weighted by $\tan\theta_{L}$, assuming the uncertainty of $dL$ to be proportional to $\cot\theta_{L}$: $c=\Sigma_{L}dL\tan\theta_{L}/\Sigma_{L}\tan\theta_{L}$.
This corresponds to a $\theta$ average of $dL$ with the weight of $\sin\theta$.
We confirmed that this method yields nearly equivalent or slightly overestimated $c$-axis lengths for $x=0.1$ samples if we use the diffraction data in ref.\cite{Fujii2002}.

In Fig. \ref{fig:c-Tc}, the relationship between $T_{\rm c}$ and $c$-axis length for the sample with $x=0.2$ is shown by filled circles.
For comparison, $x=0.1$ data from ref. \cite{Fujii2002} are also plotted with open circles.
In cuprate superconductors, there is a general tendency that the $c$-axis length decreases with increasing hole doping due to stronger Coulomb attraction between the negative oxygen ions and the in-plane hole carriers.
In keeping with this tendency, our $x=0.2$ crystal shows a qualitatively similar $T_{\rm c}$ - $c$-axis length relationship to that of the $x=0.1$ sample.
However, several differences are evident.
First, our sample has shorter $c$-axis lengths for the same annealing conditions except for the case of the condition (1), the strongest oxidation condition.
From this, we can confirm that Bi substitution with $x>0.1$ is achieved.
Second, the $T_{\rm c}$ saturation observed on the overdoped side for $x=0.1$ seems not to be realized even for the same oxidation condition.
For the $x=0.2$ sample, the $c$-axis contracted little from the oxidation conditions (2) to (1), while it contracted substantially in the $x=0.1$ sample.
This indicates that oxidation becomes more difficult on the overdoped side in the $x=0.2$ sample, possibly due to the Bi substitution.
The effect of Bi substitution in Bi2212 was studied in detail in ref. \cite{Yamashita2009} and summarized as follows: (1) the reduction of doping level is proportional to the increase in Bi content, (2) the maximum superconducting transition temperature $T_{\rm c,max}$ is reduced, (3) there is an increase in excess oxygen.
\begin{figure}[t]
	\begin{center}
	\includegraphics[width=8.5cm,clip]{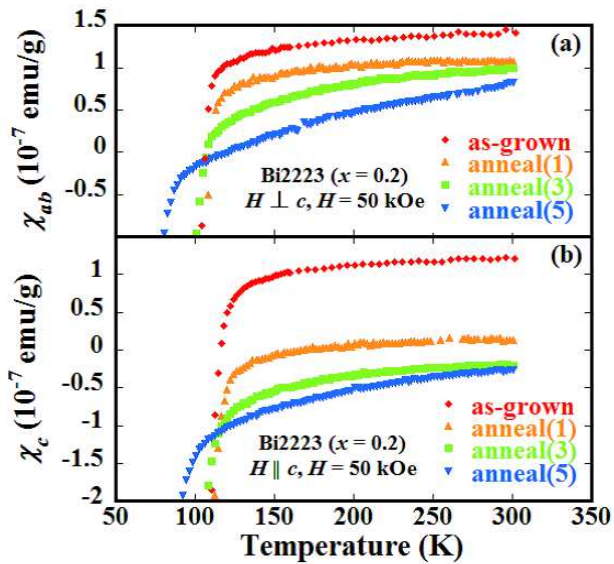}
 	\caption{(Color online) Temperature and doping dependences of anisotropic normal state susceptibilities (a)$\chi_{ab}(T)$ ($H\perp c$) and (b)$\chi_{c}(T)$ ($H\parallel c$) under a magnetic field $H=50$ kOe.}{}
 	\label{fig:susab-susc}
	\end{center}
\end{figure}
In Bi2212, the excess negative charges introduced to the plane by substituted Bi ions are partially compensated by the increase in excess oxygen and only 20 \% of substituted Bi contributes to the reduction in doping level.
Since $x=0.2$ and 0.1 samples of Bi2223 show comparable $T_{\rm c}$s for the same oxidation conditions, excess Bi is thought to be fully compensated by the additionally absorbed oxygen.
This gives a further lattice contraction along the $c$-axis as can consistently be seen in Bi2212 \cite{Yamashita2009}.
Moreover, the increase in the oxygen uptake due to substituted Bi makes further oxidation difficult on the overdoped side.
In addition, unexpectedly, $T_{\rm c,max}$ of the $x=0.2$ sample is essentially same as that of $x=0.1$ sample, while in Bi2212 about an 8 K reduction in $T_{\rm c,max}$ was reported \cite{Eisaki2004}.
This is a novel feature of the multilayered structure, where the bulk $T_{\rm c}$ is believed to be determined by the highest one among the intrinsic $T_{\rm c}$s of the nonuniformly doped individual $\rm CuO_{2}$ layers.
On the overdoped side of multilayered superconductors, $T_{\rm c}$ is thought to be determined by the optimally doped IP \cite{Fujii2002,Tokunaga2000}.
Disorder introduced by Bi$^{3+}$ ions on the Sr site could only affect the OP which does not directly govern the $T_{\rm c}$ of the sample.
Thus, in this $x=0.2$ case, the disorder might not affect the IP and $T_{\rm c,max}$ was kept unchanged.
The invariable $T_{\rm c,max}$ indicates that the interaction between IP and $\rm BiO-SrO$ layer is very weak in optimally or slightly overdoped region.

\begin{figure}[t]
	\begin{center}
	\includegraphics[width=8.5cm,clip]{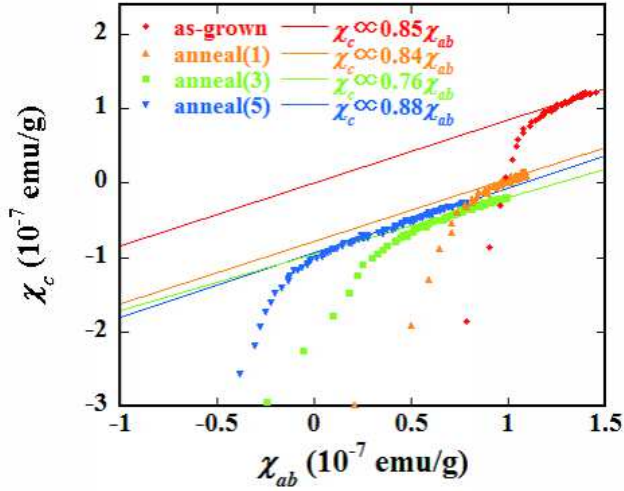}
 	\caption{(Color online) $\chi_{ab}-\chi_{c}$ plots with an implicit parameter temperature $T$ for each doping level.}{}
 	\label{fig:chiab-chic-asgrown-anneal5}
	\end{center}
\end{figure}

In Fig.\ref{fig:susab-susc}, temperature dependences of the anisotropic normal state susceptibilities $\chi_{ab}$ and $\chi_{c}$ at various doping levels are shown.
Here, we assume that these anisotropic susceptibilities are composed of five components as follows:
\begin{equation}\label{chi_alpha}
\begin{split}
\chi_{\alpha}(T) = &\chi^{\rm dia} + \chi^{\rm VV}_{\alpha} + 2\chi^{\rm spin}_{{\rm OP},\alpha}(T) \\
 		   &+\chi^{\rm spin}_{{\rm IP},\alpha}(T)+\chi^{\rm FD}_{\alpha}(T)
\end{split}
\end{equation}
where $\chi^{\rm dia}$ is the isotropic Larmor diamagnetic susceptibility, $\chi^{\rm VV}_{\alpha}$ is the anisotropic Van Vleck paramagnetic susceptibility, $\chi^{\rm spin}_{{\rm OP},\alpha}(T)$ and $\chi^{\rm spin}_{{\rm IP},\alpha}(T)$ are the anisotropic spin susceptibilities in the OP and IP, respectively, and $\chi^{\rm FD}_{\alpha}(T)$ is the anisotropic diamagnetic susceptibility due to fluctuation diamagnetism above $T_{\rm c}$.
Both $\chi^{\rm dia}$ and $\chi^{\rm VV}_{\alpha}$ are expected to be small and temperature and doping independent over the experimental temperature range.
A possible contribution to $\chi_{\alpha}$ from Landau diamagnetism is absorbed into the spin susceptibilities because it has the same temperature dependence as $\chi^{\rm spin}(T)$.
The overall magnitudes of $\chi_{ab}$ and $\chi_{c}$ monotonically decrease with decreasing doping level.
This is derived from a reduction in spin susceptibility due to the decrease in electronic density of states (DOS) near the Fermi level, as well as the opening of a pseudogap in the underdoped regime.
The absolute values and the temperature dependences of $\chi_{ab}$ for as-grown, anneal(1), anneal(3), and anneal(5) in this work are comparable to those of $\chi_{ab}$ with $x=0.1$ for corresponding annealing condition in ref. \cite{Fujii2002}.

Now let us discuss the temperature dependence of $\chi_{ab}$ and $\chi_{c}$ at each doping level.
The gradual decrease in the susceptibilities can be seen from well above $T_{\rm c}$.
This decrease should be caused by the opening of a pseudogap.
However, we did not determine the characteristic temperature $T^{*}$ of the opening of the pseudogap since $T^{*}$ is expected to become very high in the underdoped regime as in the case of Bi2212.
When temperature is further lowered to the vicinity of $T_{\rm c}$, $\chi^{\rm FD}_{\alpha}$ becomes significant.

\begin{figure}[t]
	\begin{center}
	\includegraphics[width=8.5cm,clip]{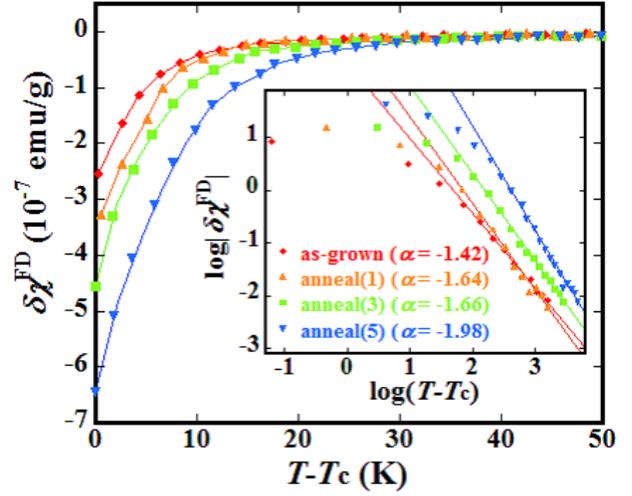}
 	\caption{(Color online) $T-T_{\rm c}$ dependence of $\delta\chi^{\rm FD}$ for each doping level, where $T_{\rm c}$s are determined by magnetization measurement applying the magnetic field 4 Oe for each. The inset shows temperature and doping dependences of $\log \lvert \delta\chi^{\rm FD}\rvert$. Each solid line is drawn by fitting the data from $T_{\rm FD}$ down to the temperature 15 K lower than $T_{\rm FD}$. The absolute value of the slope of the fitting line corresponding to the exponent $\alpha$ increases with increasing doping level.}{}
 	\label{fig:deltachiscf}
	\end{center}
\end{figure}

The fluctuation diamagnetic component can be extracted from the original data if the background normal state susceptibility is known.
However, it is apparently temperature dependent in this case and there is no reliable way to estimate it.
Here, we used the strong anisotropy in the fluctuation diamagnetism to extract it.
First, $\chi_{c}(T)$ was plotted as a function of $\chi_{ab}(T)$ with temperature $T$ as an implicit parameter, following the method described in ref. \cite{Watanabe2000, Matsuda2001}.
The plots are shown in Fig.\ref{fig:chiab-chic-asgrown-anneal5} for each doping level.
In the region far above $T_{\rm c}$, they showed linear relations with almost the same slope.
This indicates that the temperature dependent spin susceptibility has a doping and independent anisotropy ratio.
The deviation from the linear background is attributed to $\chi^{\rm FD}_{\alpha}$ which has a different anisotropy ratio from that of the spin susceptibility.
Based on the simple Gaussian superconducting fluctuation theory\cite{Tinkham}, anisotropy in $\chi^{\rm FD}_{\alpha}$ comes from anisotropy in the effective shape of fluctuating domain whose dimension is proportional to $\xi_{ab}\times\xi_{ab}\times\xi_{c}$, where $\xi_{ab}$ and $\xi_{c}$ are the in-plane and out-of-plane coherence lengths respectively.
$\chi^{\rm FD}_{\alpha}$ can be estimated as the vertical deviation from the linear background assuming $\chi^{\rm spin}_{{\rm OP},\alpha}(T)=(g^{\rm OP}_{\alpha})^{2}\chi^{\rm spin}_{\rm OP}(T), \chi^{\rm spin}_{{\rm IP},\alpha}(T)=(g^{\rm IP}_{\alpha})^{2}\chi^{\rm spin}_{\rm IP}(T)$, $g^{\rm IP}_{\alpha}=\gamma g^{\rm OP}_{\alpha}$, where $\gamma$ is a constant and $g_{\alpha}$ is a gyromagnetic ratio of the conduction electrons, and that $g^{\rm IP}_{\alpha}$ and $g^{\rm OP}_{\alpha}$ are temperature and doping independent.
Using eq.(1) and the above assumptions, the fluctuation diamagnetic component $\delta\chi^{\rm FD}$ is calculated to be
\begin{equation}
\begin{split}
\delta\chi^{\rm FD}(T)&=\chi_{c}(T)-\left(\frac{g^{\rm OP}_{c}}{g^{\rm OP}_{ab}}\right)^{2}\chi_{ab}(T)-\chi_{0}\\
		   &=\chi^{\rm FD}_{c}(T)-\left(\frac{g^{\rm OP}_{c}}{g^{\rm OP}_{ab}}\right)^{2}\chi^{\rm FD}_{ab}(T) \\
&\sim\chi^{\rm FD}_{c}(T)
\end{split}
\end{equation}
where $(g^{\rm OP}_{c}/g^{\rm OP}_{ab})^{2}$ and $\chi_{0}\equiv\{1-(g^{\rm OP}_{c}/g^{\rm OP}_{ab})^{2}\}\chi^{\rm dia}+\chi^{\rm VV}_{c}-(g^{\rm OP}_{c}/g^{\rm OP}_{ab})^{2}\chi^{\rm VV}_{ab}$ correspond to the slope and intercept of the linear part of the $\chi_{ab}-\chi_{c}$ plot respectively.
Since $\chi^{\rm FD}_{ab}$ is thought to be much smaller than $\chi^{\rm FD}_{c}$ and $(g^{\rm OP}_{c}/g^{\rm OP}_{ab})^{2}$ is less than 1 (see Fig.\ref{fig:chiab-chic-asgrown-anneal5}), $\delta\chi^{\rm FD}$ is nearly equal to $\chi^{\rm FD}_{c}$.
Thus, $\chi^{\rm FD}_{c}$ can be estimated under the above assumptions.

\begin{figure}[t]
	\begin{center}
	\includegraphics[width=8.5cm,clip]{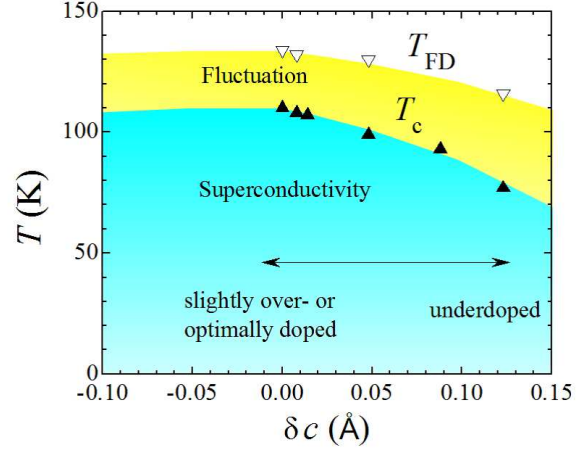}
 	\caption{(Color online) The superconducting temperature $T_{\rm c}$s and the onset of fluctuation diamagnetism temperature $T_{\rm FD}$s are shown with filled triangles and open inverted triangles respectively.
$T_{\rm FD}$s show the dome shaped doping dependence as well as $T_{\rm c}$s.
$\delta c$ is the variation in the $c$-axis length from that of the as-grown sample.
Since the $T_{\rm FD}$ curve highly resembles the $T_{\rm onset}$ curve in the phase diagram of Bi2212 reported by Wang et al.\cite{Wang2005}, the same physical property might have been detected in our DC susceptibility measurements and their Nernst experiments.
}{}
 	\label{fig:Tc-Tscf}
	\end{center}
\end{figure}

$T-T_{\rm c}$ dependences of $\delta\chi^{\rm FD}$ for various doping levels are shown in Fig.\ref{fig:deltachiscf}.
Here, we used $T_{\rm c}$s as the superconducting transition temperature determined by magnetization measurements under a magnetic field of 4 Oe.
$\delta\chi^{\rm FD}$s increase divergently toward $T_{\rm c}$.
We estimated the characteristic temprature $T_{\rm FD}$ at which $\delta\chi^{\rm FD}$ becomes 2 \% of the full value at $T_{\rm c}$ for anneal(5).
Obtained $T_{\rm FD}$s were 134 K, 132 K, 130 K, and 116 K for samples that were measured as-grown and annealing under the conditions (1), (3), and (5) respectively, and they are plotted against $\delta c$ (the variation in the $c$-axis length from that of the as-grown sample) together with corresponding $T_{\rm c}$s 110 K, 108 K, 99 K, and 77 K respectively in Fig.\ref{fig:Tc-Tscf}.
For each doping level, $T_{\rm FD}$ was about 30 K higher than $T_{\rm c}$, and these results are similar to the experimental results of Bi2212 reported by Wang \textit{et al}\cite{Wang2005} demonstrating that $\delta\chi^{\rm FD}$ is a fluctuation diamagnetic component.
$\delta\chi^{\rm FD}$ also increases with decreasing doping.
This behavior can be interpretted as that the difference between $T_{\rm c}$ and $T_{\rm FD}$ increases with decreasing doping.
$\chi^{\rm FD}_{c}$ is expressed as functions of doping level $p$ and temperature $T$ as follows:
\begin{equation}
\begin{split}
\chi^{\rm FD}_{c}(p,T)\propto \frac{k_{\rm B}T}{V(p,T)}\xi^{2}_{ab}(p,T)\langle r^{2}\rangle _{\rm eff}(p,T)
\end{split}
\end{equation}
where $V\sim\pi\xi^{2}_{ab}\cdot{\rm max}(\xi_{c}(p,T),d)$ is the coherence volume, $d$ is the distance between Cu-O blocks, and $\langle r^{2}\rangle _{\rm eff}$ is the mean-square radius of the fluctuating domain, which is roughly represented as $\langle r^{2}\rangle _{eff}\sim\xi^{2}_{ab}$ for $H\parallel c$ in this case.
Consequently, $\chi^{\rm FD}_{c}$ is calculated to be
\begin{equation}
\begin{split}
\chi^{\rm FD}_{c}(p,T)\propto k_{\rm B}T\frac{\xi^{2}_{ab}(p,T)}{{\rm max}(\xi_{c}(p,T),d)}.
\end{split}
\end{equation}

$\delta\chi^{\rm FD}$ is plotted as a function of $\ln(T-T_{\rm c})$ as can be seen in the inset of Fig.\ref{fig:deltachiscf}.
$\delta\chi^{\rm FD}$ exhibits a divergent temperature dependence $(T-T_{\rm c})^{\alpha}$ with a doping dependent exponent $\alpha$ varing from -1.4 to -2.0.
In addition, the exponents $\alpha$ determined by our experiments are rather lower than those expected in a classical 2D superconductor ($\alpha=-1$) \cite{Tinkham}.
The same analysis on Bi2212 gave an essentially doping independent exponent $\alpha\sim-2.3$ \cite{Matsuda2001}, lower than that for any doping of Bi2223.
Recent Nernst experiments for cuprates indicates that unbound vortex-antivortex pairs exist well above $T_{\rm c}$ and they disorder the long-range phase coherence \cite{Li2010}.
In such vortex liquid state, temperature dependence of the correlation length $\xi$ is expressed as 
\begin{equation}
\begin{split}
\xi '_{\alpha}&\propto\exp\left[\left(\frac{B}{T-T_{\rm c}}\right)^{1/2}\right]
\end{split}
\end{equation}
where $B$ is a constant.\cite{Halperin1979}
Since the exponential temperature dependence of $\xi '_{\alpha}$ is stronger than any other power law temperature dependence of the form of $(T-T_{c})^{\alpha}$, the nominal exponent will be lifted up if $\xi_{\alpha}$ includes some component of $\xi '_{\alpha}$.
The disagreement between the observed exponents of Bi2223 or Bi2212 and the expected one in simple Gaussian fluctuation theory implies that both Bi2223 and Bi2212 includes vortex liquid region above $T_{\rm c}$.
The decrease in $\left|\alpha\right|$ of Bi2223 on doping suggests decrease of the vortex liquid domain in the sample and/or that the system becomes even less two-dimensional.
This behavior is associated with a more rapid weakening of the interlayer coupling of Bi2223 on underdoping than that of Bi2212.
The deviation of the plots from fitting line near $T_{\rm c}$ may result from the fact that the temperature dependence of $\chi^{\rm FD}_{c}$ becomes weak since ${\rm max}(\xi_{c},d)$ changed from $d$ to $\xi_{c}$ due to divergence of $\xi_{c}$ on cooling down toward  $T_{\rm c}$.

One possible origin of such modification is the existence of the IP in Bi2223.
In this system, nonuniformly doped $\rm CuO_{2}$ layers were confirmed by ARPES measurements, which estimated the doping level of the IP to be 7 \%, while that of OP was 23 \% in optimally doped samples\cite{Ideta2009}.
According to the ARPES result, slight underdoping can easily drive an IP insulating or possibly into an AF ordered state.
The insulating IP disrupts superconducting coherence among the three layers (OP-IP-OP) and decouples the system into electromagnetically coupled OPs in the underdoped regime.

\section{Conclusions}
High quality single crystals of ${\rm Bi}_{2+x}{\rm Sr}_{2-x}{\rm Ca}_{2}{\rm Cu}_{3}{\rm O}_{10+\delta}$ (Bi2223) with $x=0.2$ were grown using a traveling solvent floating zone method in order to investigate dimensionality of superconductivity in highly underdoped Bi2223 crystals.
Obtained crystals were characterized by X-ray diffraction, DC susceptibility and resistivity measurements, confirming Bi2223 to be the main phase.
The highest quality crystal was annealed under various oxygen partial pressures, adjusting its carrier density from slightly underdoped to highly underdoped.
From the relationship between $T_{\rm c}$ and $c$-axis length, we could conclude that higher Bi substitution for Sr ($x>0.1$) is successfully achieved.
However, Bi2223 with $x=0.2$ did not show a decrease in $T_{\rm c,max}$, compared with $x=0.1$ Bi2223.
This behavior confirms that $T_{\rm c}$ of the inner layer limits the bulk $T_{\rm c}$ in the optimally-doped region.
The fluctuation diamagnetic component was extracted from the anisotropic normal state susceptibilities $\chi_{ab}(T)$ ($H\perp c$) and $\chi_{c}(T)$ ($H\parallel c$).
The temperature dependence of this component became strong with underdoping, suggesting a reduction of the superconducting dimensionality and/or an increase of vortex liquid domain.
This behavior supports the view that inner $\rm CuO_{2}$ layer which is relatively underdoped compared to outer layers disrupts superconducting coherence among the three layers and change interlayer coupling situation more largely than that of Bi2212.

\section*{Acknowledgment}
We thank T. Fujii for providing us law XRD data and D. C. Peets for valuable discussions. This work was supported in part by "Academic Frontier" project from MEXT and grants from the Ministry of Education, Culture and Science of Japan.


\begin{thebibliography}{99} 
\bibitem{Iyo2007} A. Iyo, Y. Tanaka, H. Kito, Y. Kodama, P. M. Shirage, D. D. Shivagan, H. Matsuhata, K. Tokiwa, and T. Watanabe: J. Phys. Soc. Jpn. \textbf{76} (2007) 094711.
\bibitem{Chakravarty2004} S. Chakravarty, H. Kee, and K. V\"{o}lker: Nature \textbf{428} (2004) 53.
\bibitem{Trokiner1991} A. Trokiner, L. Le Noc, J. Schneck, A. M. Pougnet, R. Mellet, J. Primot, H. Savary, Y. M. Gao, and S. Aubry: Phys. Rev. B \textbf{44} (1991) 2426.
\bibitem{Statt1993} B. W. Statt and L. M. Song: Phys. Rev. B \textbf{48} (1993) 3536.
\bibitem{Kontos1998} A. G. Kontos and R. Dupree: Physica C \textbf{317-318} (1998) 565.
\bibitem{Piskunov1998} Y. V. Piskunov, K. N. Mikhalev, Yu. I. Zhdanov, A. P. Gerashenko, S. V. Verkhovskii, K. A. Okulova, E. Yu. Medvedev, A. Yu. Yakubovskii, L. D. Shustov, P. V. Bellot, and A. Trokiner : Physica C \textbf{300} (1998) 225.
\bibitem{Tokunaga1999} Y. Tokunaga, H. Kotegawa, K. Ishida, G. -q. Zheng, Y. Kitaoka, K. Asayama, K. Tokiwa, A. Iyo, and H. Ihara: J. Low Temp. Phys. \textbf{117} (1999) 473.
\bibitem{Kotegawa2001} H. Kotegawa, Y. Tokunaga, K. Ishida, G. -q. Zheng, Y. Kitaoka, H. Kito, A. Iyo, K. Tokiwa, T. Watanabe, and H. Ihara: J. Phys. and Chem. of Solids \textbf{62} (2001) 171-175.
\bibitem{Ideta2009} S. Ideta, K. Takashima, M. Hashimoto, T. Yoshida, A. Fujimori, H. Anzai, T. Fujita, Y. Nakashima, A. Ino, M. Arita, H. Namatame, M. Taniguchi, K. Ono, M. Kubota, D. H. Lu, Z.-X. Shen, K. M. Kojima, and S. Uchida: Phys. Rev. Lett. \textbf{104} (2010) 227001.
\bibitem{Chen2009} Y. Chen, A. Iyo, W. Yang, A. Ino, S. Johnston, H. Eisaki, H. Namatame, M. Taniguchi, T. P. Devereaux, Z. Hussain, and Z. -X. Shen: Phys. Rev. Lett. \textbf{103} (2009) 036403.
\bibitem{Kotegawa2004} H. Kotegawa, Y. Tokunaga, Y. Araki, G. -q. Zheng, Y. Kitaoka, K. Tokiwa, K. Ito, T. Watanabe, A. Iyo, Y. Tanaka, and H. Ihara: Phys. Rev. B \textbf{69} (2004) 014501.
\bibitem{Mukuda2006} H. Mukuda, M. Abe, Y. Araki, Y. Kitaoka, K. Tokiwa, T. Watanabe, A. Iyo, H. Kito, and Y. Tanaka: Phys. Rev. Lett. \textbf{96} (2006) 087001.
\bibitem{Kitaoka2007} Y. Kitaoka, H. Mukuda, S. Shimizu, M. Abe, A. Iyo, Y. Tanaka, H. Kito, K. Tokiwa, and T. Watanabe: J. Mag. Mag. Mat. \textbf{310} (2007) 467-473.
\bibitem{Shimizu2009} S. Shimizu, T. Sakaguchi, H. Mukuda, Y. Kitaoka, P. M. Shirage, Y. Kodama, and A. Iyo: Phys. Rev. B \textbf{79} (2009) 064505.
\bibitem{Mori2005} M. Mori and S. Maekawa: Phys. Rev. Lett. \textbf{94} (2005) 137003.
\bibitem{Piriou2008} A. Piriou, Y. Fasano, E. Giannini, and \O. Fischer: Phys. Rev. B \textbf{77} (2008) 184508.
\bibitem{Fujii2001} T. Fujii, T. Watanabe, and A. Matsuda: J. Cryst. Growth \textbf{223} (2001) 175.
\bibitem{Liang2002} B. Liang, C. T. Lin, P. Shang, and G. Yang: Physica C \textbf{383} (2002) 75.
\bibitem{Giannini2004} E. Giannini, V. Garnier, R. Gladyshevskii, and R. Fl\"ukiger: Supercond. Sci. Technol. \textbf{17} (2004) 220-226.
\bibitem{Tokiwa1998} K. Tokiwa, Y. Tanaka, A. Iyo, Y. Tsubaki, K. Tanaka, J. Akimoto, Y. Oosawa, N. Terada, M. Hirabayashi, M. Tokumoto, S. K. Agarwal, T. Tsukamoto, and H. Ihara: Physica C \textbf{298} (1998) 209-216.
\bibitem{Iyo2004} A. Iyo, M. Hirai, K. Tokiwa, T. Watanabe, and Y. Tanaka: Supercond. Sci. Technol. \textbf{17} (2004) 143-147.
\bibitem{Li2010} L. Li, Y. Wang, S. Komiya, S. Ono, Y. Ando, G. D. Gu, and N. P. Ong: Phys. Rev. B \textbf{81} (2010) 054510.
\bibitem{Watanabe1997} T. Watanabe, T. Fujii, and A. Matsuda: Phys. Rev. Lett. \textbf{79} (1997) 2113.
\bibitem{Yamashita2009} S. Yamashita, T. Kasai, T. Fujii, T. Watanabe, and A. Matsuda: to be published in Physica C.
\bibitem{Kulakov2006} A. B. Kulakov, D. Maier, A. Maljuk, I. K. Bdikin, and C. T. Lin: J. Cryst. Growth \textbf{296} (2006) 69.
\bibitem{Fujii2002} T. Fujii, I. Terasaki, T. Watanabe, and A. Matsuda: Phys. Rev. B \textbf{66} (2002) 024507.
\bibitem{Eisaki2004} H. Eisaki, N. Kaneko, D. L. Feng, A. Damascelli, P. K. Mang, K. M. Shen, Z. -X. Shen, and M. Greven: Phys. Rev. B \textbf{69} (2004) 064512.
\bibitem{Tokunaga2000} Y. Tokunaga, K. Ishida, Y. Kitaoka, K. Asayama, K. Tokiwa, A. Iyo, and H. Ihara: Phys. Rev. B \textbf{61} (2000) 9707.
\bibitem{Watanabe2000} T. Watanabe, T. Fujii, and A. Matsuda: Phys. Rev. Lett. \textbf{84} (2000) 5848.
\bibitem{Matsuda2001} A. Matsuda, S. Sugita, T. Fujii, and T. Watanabe: J. Phys. and Chem. of Solids \textbf{62} (2001) 65-68.
\bibitem{Tinkham} M. Tinkham: \textit{Introduction to Superconductivity} (Dover, New York, 2004) 2nd ed., p. 307. 
\bibitem{Wang2005} Y. Wang, L. Li, M. J. Naughton, G. D. Gu, S. Uchida, and N. P. Ong: Phys. Rev. Lett. \textbf{95} (2005) 247002.
\bibitem{Halperin1979} B. I. Halperin and D. R. Nelson: J. Low Temp. Phys. \textbf{36} (1979) 599.
\end{thebibliography}
\end{document}